# Asymmetric Bias Dependence in Double Spin Filter Tunnel Junctions

G. X. Miao* and J. S. Moodera

Francis Bitter Magnet Laboratory, Massachusetts Institute of Technology, Cambridge,

MA 02139

In double spin filter (SF) tunnel junctions, the spin information is generated and analyzed purely from the SF effect with nonmagnetic electrodes. In this article we numerically evaluate the bias dependence of magnetoresistance in such tunnel junctions (nonmagnetic metal / SF / nonmagnetic insulator / SF / nonmagnetic metal), particularly in cases when different SF materials are utilized. A large magnetoresistance with nonmonotonic and asymmetric bias dependence is expected within the frame of WKB approximation. We illustrate the systematic influence of tunnel barrier height, tunnel barrier thickness, and exchange energy splitting on magnetoresistance, particularly focusing on the asymmetric behavior of the magnetoresistance bias dependence.

Spin filtering (SF) is a unique approach to generate highly spin polarized tunnel currents and is invaluable in the fast evolving field of spintronics [1]. SF takes advantage of the fact that in magnetic insulator/semiconductors, the conduction band is spontaneously exchange split, thus the electrons' tunneling through barriers formed with these materials is spin orientation sensitive. Due to the exponential dependence of tunneling probability on tunnel barrier height, spin carriers facing the higher barrier are effectively filtered out and the net tunnel current tends towards completely spin polarized [2-4]. It can be seen that the SF effect can be readily applied for spin generation and spin detection. When two such SF barriers are combined together, i.e., as double spin filter tunnel junction [5], the total tunnel conductance is dependent on the relative magnetic alignment between the two layers, and very large tunnel magnetoresistance (TMR) is expected. Experimentally, both quasi SF tunnel junctions with single SF barrier [6-13] and full SF tunnel junctions with double SF barriers [14] were successfully demonstrated. There has been strong activity in this field with many interesting SF candidates: for example, the ferrites ($CoFe_2O_4$, $NiFe_2O_4$ etc.) have shown some effect at room temperature [7-10]; whereas at low temperatures the Eu chalcogenides (EuO, EuS, EuSe etc.) showed the highest SF efficiency experimentally [2-4,6,11,14]. Furthermore, the perovskite family ($BiMnO_3$, $La_{0.1}Bi_{0.9}MnO_4$ etc.) also exhibited multiferroic properties given the right level of doping [12,13].

The bias dependence of TMR in SF tunnel junctions is highly nonmonotonic with a pronounced MR peak [11,14,15], due to the establishment of the Fowler-Nordheim (FN) tunneling [16]. In previous theoretical reports [5,15,17], the effects of SF layer thickness and bias voltage on MR are well described, and in these models the two SF layers are constructed to be identical. On the other hand, one very important situation in which asymmetric SF barriers are utilized has largely been overlooked in literature, despite such structures are likely the most common situation for separating magnetic coercivities. In our previous study [14], we have shown experimentally that TMR is very asymmetric versus bias voltage polarity when the two SF barriers consist of the same material but have different thicknesses. In this article, we aim to numerically evaluate the expected bias dependence in double SF tunnel junctions within the framework of Wentzel-Kramers-Brillouin (WKB) approximations, especially in the cases when SFs with different barrier height, exchange splitting energy, and thickness, are combined. The overall behavior predicted with the simulations could be pertinent for future studies on double SF tunnel junctions.

Before going into the numerical calculations for double SF tunnel junctions, we first start with the dependence of SF efficiency on the barrier parameters. For a single SF barrier with average barrier height $\varphi$ (i.e., the barrier height above Curie temperature), exchange splitting $2\Delta$, and thickness $t$, the tunnel barrier height is $\varphi - \Delta$ for the spin-up electrons, and $\varphi + \Delta$ for the spin-down electrons, respectively. The ratio between the spin-up and -down electron conductance is,

$$\frac{G_\uparrow}{G_\downarrow} = \exp\left[-2t\left(\sqrt{\frac{2m_e}{\hbar^2}(\varphi-\Delta)} - \sqrt{\frac{2m_e}{\hbar^2}(\varphi+\Delta)}\right)\right] \xrightarrow{\Delta \ll \varphi} \exp\left[4\sqrt{\frac{2m_e}{\hbar^2}}\frac{\Delta}{\sqrt{\varphi}}t\right] \quad (1)$$

Here $m_e$ is the free electron mass. The above relation gives the general guidance for achieving high SF efficiency: thicker SF layer with larger exchange splitting (relative to the average barrier height) is beneficial, and the improvement is expected to be dramatic due to their exponential dependences.

In Fig.1a we illustrate the model system we are treating. It consists of nonmagnetic metals as electrodes, and the tunnel barrier is formed with two SF layers plus an additional nonmagnetic insulator layer in between. The spacer layer is included in our calculation to magnetically decouple the two ferromagnetic SFs [18], and we will see from the calculations that its effect on the final TMR is negligible. For simplicity, the electric field distribution is assumed to be uniform inside the barrier region. Hence the tunnel barriers are distorted into trapezoidal shapes under biases, and the barrier height as a function of the position $x$ can be mathematically expressed as follows (see Fig.1b and c for the corresponding barriers in the P and AP states),

$$\varphi^P_{\uparrow,\downarrow}(x) = \begin{cases} \varphi_1 \mp \Delta_1 + \frac{eV}{d}(x-d), & (0<x<t_1) \\ \varphi_{spacer} + \frac{eV}{d}(x-d), & (t_1<x<d-t_2) \\ \varphi_2 \mp \Delta_2 + \frac{eV}{d}(x-d), & (d-t_2<x<d) \end{cases} \quad \varphi^{AP}_{\uparrow,\downarrow}(x) = \begin{cases} \varphi_1 \mp \Delta_1 + \frac{eV}{d}(x-d), & (0<x<t_1) \\ \varphi_{spacer} + \frac{eV}{d}(x-d), & (t_1<x<d-t_2) \\ \varphi_2 \pm \Delta_2 + \frac{eV}{d}(x-d), & (d-t_2<x<d) \end{cases}.$$

Here the subscripts 1 and 2 denote SF1 and SF2, $\varphi_{spacer}$ is the barrier height of the spacer layer, $d = t_1 + t_2 + t_{spacer}$ is the total barrier thickness, and $V$ is the applied bias voltage on electrode1 relative to electrode2. In WKB approximations, the transmission probability $T$ for an electron with energy $E$ can be expressed as follows,

$$T_{\uparrow,\downarrow}(E) \cong \exp\left[-2\int_0^d \sqrt{\frac{2m_e}{\hbar^2}(\varphi_{\uparrow,\downarrow}(x)-E)}dx\right]$$

In this study only electrons incident normal to the barrier (i.e., $k_{//}=0$) were taken into account due to the large barrier thickness involved. The total dc conductance $G$ at bias voltage $V$ is an integration of all the available states above the Fermi level, and can be expressed as follows,

$$\begin{cases} G^P \propto \int_{-\infty}^{\infty} [N_{1\uparrow}(E-eV)N_{2\uparrow}(E)T_{\uparrow}^P(E) + N_{1\downarrow}(E-eV)N_{2\downarrow}(E)T_{\downarrow}^P(E)][f(E-eV)-f(E)]dE \\ G^{AP} \propto \int_{-\infty}^{\infty} [N_{1\uparrow}(E-eV)N_{2\downarrow}(E)T_{\uparrow}^{AP}(E) + N_{1\downarrow}(E-eV)N_{2\uparrow}(E)T_{\downarrow}^{AP}(E)][f(E-eV)-f(E)]dE \end{cases}$$

Here $f(E)$ is the Fermi distribution function and reduces to a step function at $T=0$; $N_1$ and $N_2$ are the DOS of the electrodes on the two sides. Here we are dealing with nonmagnetic electrodes, and we further ignore the detailed features of their conduction bands by assuming that the DOS is independent of $E$. Thus $N_{i\uparrow} = N_{i\downarrow} = n_i (i=1,2)$, where $n_i$ is a constant and can be taken out of the integral. The conductance ratio between the P and AP states ($\frac{G^P}{G^{AP}} \sim TMR+1$) is thus reduced to (at $T=0$),

$$\frac{G^P}{G^{AP}} = \frac{n_1 n_2 \int_{-\infty}^{\infty} [T_{\uparrow}^P(E) + T_{\downarrow}^P(E)][f(E-eV)-f(E)]dE}{n_1 n_2 \int_{-\infty}^{\infty} [T_{\uparrow}^{AP}(E) + T_{\downarrow}^{AP}(E)][f(E-eV)-f(E)]dE} = \frac{\int_0^{eV} [T_{\uparrow}^P(E) + T_{\downarrow}^P(E)]dE}{\int_0^{eV} [T_{\uparrow}^{AP}(E) + T_{\downarrow}^{AP}(E)]dE}.$$

For each bias voltage $V$, the calculation is simply the ratio between two double integrals, which are evaluated in Matlab by summing over a mesh with even spacing of 0.001 nm for $dx$ and $V/1000$ for $dV$, respectively. Since the tunneling electrons are not phase coherent, we only considered their decay component in this integration (i.e., only in regions where $E < \varphi(x)$). The default parameters are set as follows: the average barrier heights $\varphi_1 = \varphi_2 = 0.8$ eV, $\varphi_{spacer} = 1.5$ eV; the half exchange splitting $\Delta_1 = \Delta_2 = 0.1$ eV; and the barrier thicknesses $t_1 = t_2 = 2.5$ nm, $t_{spacer} = 1$ nm. The SF thickness of 2.5 nm is chosen to reflect the typical experimental value: it cannot be too thin as to weaken the magnetic properties, or too thick to render the junction impedance way too high. The chosen barrier heights are typical values for EuS and $Al_2O_3$ based systems. We then proceed to systematically vary the parameters and show their effects on the TMR.

Fig.2 shows the calculated conductance ratio when the exchange splitting of SF1 is fixed at $2\Delta_1 = 0.2$ eV, while that of SF2 is varied over a broad range. The expected TMR is very large in these double SF tunnel junctions, and can readily reach tens of thousands percent for a standard system. The general trend is clear, for higher exchange splitting values, the overall TMR of the system is also higher. This is because larger exchange splitting creates more dramatic difference between the decay wave vectors of spin-up and -down electrons, leading to exponential dependence of spin filtering efficiency on exchange splitting (Equ.1). Another trend is that TMR is nonmonotonic with respect to bias voltage: it has a pronounced peak on each bias polarity. This has been confirmed experimentally and attributed to the presence of FN tunneling [11,14]. Now we pay our attention to the relative amplitude and position of the MR peaks, which are very asymmetric for the two bias polarities. In our definition, positive bias will lower the Fermi level of electrode1 so that electrons first travel across SF2 to become polarized then across SF1 to be analyzed, and for negative bias it is the other way. From Fig.2 we can see that when the electrons travel from a more exchange split SF into a less split one, the TMR peak is higher than for the opposite current flow direction. This shows, as one might expect, that a stronger spin polarizing power is desired in order to achieve maximum TMR, which is suitable for applications that require maximum signal output. On the other hand, when the electrons travel in the opposite direction (i.e., from the less split SF into the more split one), TMR is relatively lower and has a flatter bias dependence, which could also be useful for certain applications where signal stability is more important. When a tunneling electron coming from the polarizer side has energy higher than the SF conduction band level on the analyzer side, it will directly enter the conduction bands and roll across under the strong electric field, and will not contribute to junction conductance change. Because these high energy electrons are exponentially more favored in the tunneling process, the total TMR drops to nearly zero at very high biases. On the other hand, the low energy electrons still cross the barriers in a spin dependent manner despite their weak contribution to the total conductance, and are responsible for the small yet nonvanishing TMR tail extending into high biases.

In Fig.3 top panel one again observes the highly asymmetric bias dependence when the average barrier height of one SF is different from the other. The general trend is that higher barrier heights lead to reduced overall TMR, because the exchange splitting is less effective and thus reduces the SF efficiency. The position of the TMR peak shifts systematically towards lower voltages for lower tunnel barriers, because FN tunneling can be established at lower biases for lower barriers. On the other hand, the maximum TMR occurs when the electrons travel from the higher SF barrier into the lower one, i.e., from the one with lower SF efficiency into the one with higher. This is opposite to what we saw earlier. The calculation reveals the origin of such a difference: in a double SF tunnel junction, SF with higher average barrier height contributes more to the total junction resistance, therefore more influential to the final TMR. Increasing the exchange splitting of the other SF can to a large extent counteract the influence, and enhance the TMR of the other bias polarity (Fig.3, bottom panel).

Next we discuss the influence of the SF layer thickness. For a given system, the SF barrier height and exchange splitting are not readily tunable, yet the barrier thickness can be easily varied in experiments. It was previously shown that thicker SF layers will lead to larger TMR because of the enhancement in SF efficiency [5]. In our calculations (Fig.4), we maintain the total barrier thickness constant and only vary the relative thickness between SF1 and SF2. The overall TMR is optimized when the two SFs are balanced in thickness, especially in the low bias region. The maximum TMR occurs when electrons travel from thicker SF into thinner one, and TMR in the opposite bias polarity is quite flat [14]. The freedom to choose barrier thickness adds further flexibility in tuning the TMR bias dependence towards desired applications.

Finally, we briefly discuss the effect of the spacer layer. As one can see from the calculations (Fig.5), the spacer layer has negligible influence on the final TMR because it does not distinguish spin-up and -down electrons. The zero bias TMR is completely independent on the spacer layer property, and the slight difference seen at high biases results from the change of voltage distribution among the layers.

In conclusion, we have performed numerical evaluations on the expected TMR bias dependence in double SF tunnel junctions. The main features are summarized below: (i) extremely high TMR can be generated in such junctions; (ii) SF efficiency and TMR can be improved by increasing the SF thickness and exchange splitting, or reducing the average barrier height; (iii) TMR is asymmetric with respect to bias voltage and has a pronounced peak in each polarity; (iv) the maximum TMR is achieved when the electrons start from a more efficient SF, i.e., from a thicker/more exchange split SF into a thinner/less exchange split one; (v) the insulating spacer layer has negligible influence on TMR as well as its bias dependence.

This work is supported by NSF, DARPA, ONR and KIST-MIT project funds.

E-mail: gxmiao@mit.edu

1. J.S. Moodera, T.S. Santos, T. Nagahama, J. Phys. Cond. Mat. **19**, 165202 (2007).

2. J.S. Moodera, X. Hao, G.A. Gibson, R.T. Meservey, Phys. Rev. Lett. **61**, 637 (1988).

3. T.S. Santos, J.S. Moodera, Phys. Rev. B **69**, 241203(R) (2004).

4. T.S. Santos, J.S. Moodera, K.V. Raman, E. Negusse, J. Holroyd, J. Dvorak, M. Liberati, Y.U. Idzerda, E. Arenholz, Phys. Rev. Lett. **101**, 147201 (2008).

5. D.C. Worledge, T.H. Geballe, J. Appl. Phys. **88**, 5277 (2000).

6. P. LeClair, J.K. Ha, H.J.M. Swagten, J.T. Kohlhepp, C.H. van de Vin, W.J.M. de Jonge, Appl. Phys. Lett. **80**, 625 (2002).

7. M.G. Chapline, S.X. Wang, Phys. Rev. B **74**, 014418 (2006).

8. U. Lüders, M. Bibes, K. Bouzehouane, E. Jacquet, J.-P. Contour, S. Fusil, Appl. Phys. Lett. **88**, 082505 (2006).

9. U. Lüders, M. Bibes, S. Fusil, K. Bouzehouane, E. Jacquet, C.B. Sommers, J.-P. Contour, J.-F. Bobo, A. Barthélémy, A. Fert, P.M. Levy, Phys. Rev. B **76**, 134412 (2007).

10. A.V. Ramos, M.-J. Guittet, J.-B. Moussy, R. Mattana, C. Deranlot, F. Petroff, C. Gatel, Appl. Phys. Lett. **91**, 122107 (2007).

11. T. Nagahama, T.S. Santos, J.S. Moodera, Phys. Rev. Lett. **99**, 016602 (2007).

12. M. Gajek, M. Bibes, A. Barthélémy, K. Bouzehouane, S. Fusil, M. Varela, J. Fontcuberta, A. Fert, Phys. Rev. B **72**, 020406(R) (2005).

13. M. Gajek, M. Bibes, S. Fusil, K. Bouzehouane, J. Fontcuberta, A. Barthélémy, A. Fert, Nat. Mater. **6**, 296 (2007).

14. G.X. Miao, M. Müller, J.S. Moodera, Phys. Rev. Lett. (in print).

15. Z. W. Xie, B. Z. Li, J. Appl. Phys. **93**, 9111 (2003).

16. R. H. Fowler, L. Nordheim, Proc. R. Soc. A **119**, 173 (1928).

17. A. Saffarzadeh, J. Phys. Condens. Matter **15**, 3041 (2003).

18. G.X. Miao, J.S. Moodera, Appl. Phys. Lett. **94**, 18xxxx (in print, 2009).

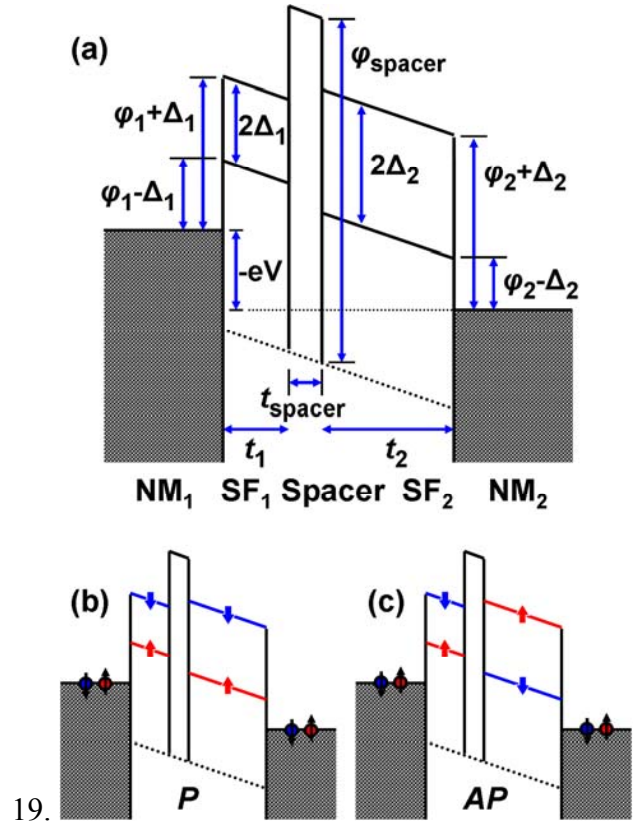

19.

Fig.1 (a) Illustration of the double SF tunnel junction structure labeling the variables used in the text. Here the bias voltage is applied on electrode1 relative to electrode2, note that a negative voltage ($V<0$) on electrode1 will actually raise its electron Fermi level. The tunnel barrier heights for spin-up and -down electrons are indicated with red and blue for spin-P (b) and -AP (c) states, respectively.

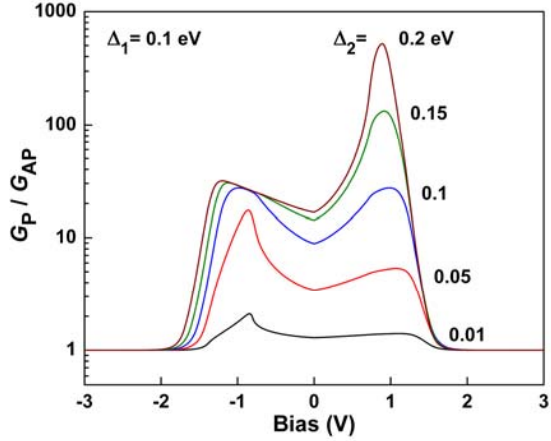

Fig.2 Bias dependence when the half exchange splitting of SF2 ($\Delta_2$) is varied from 0.01, 0.05, 0.1, 0.15, to 0.2 eV.

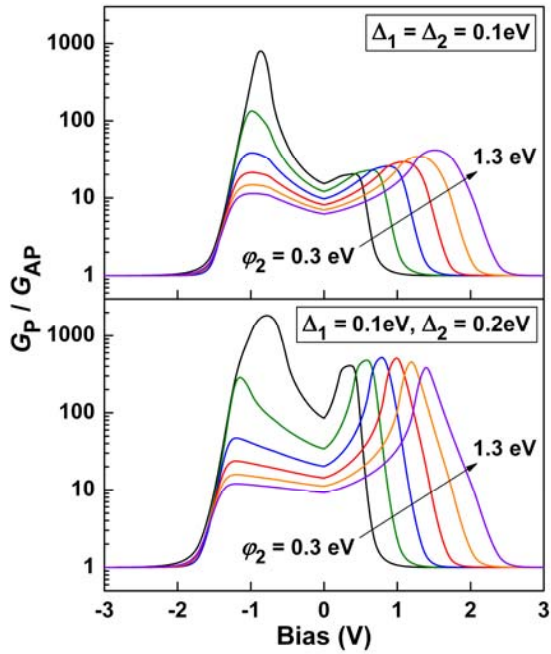

Fig.3 Bias dependence when the average barrier height of SF2 ($\varphi_2$) is systematically varied from 0.3, 0.5, 0.7, 0.9, 1.1, to 1.3 eV. The top and bottom plots illustrate the case when the exchange splittings of SF1 and SF2 are the same, and the case when they are different, respectively.

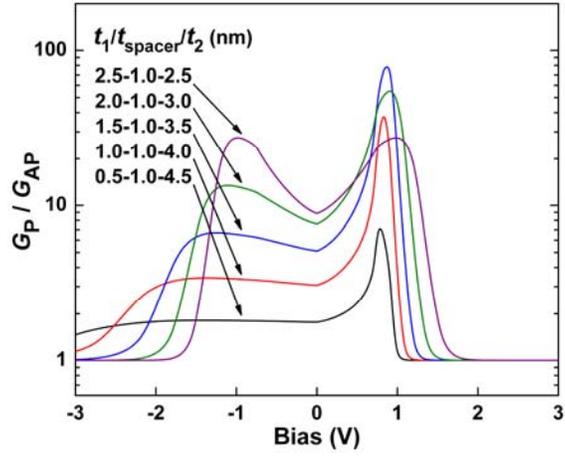

Fig.4 Bias dependence when the SF1 and SF2 thicknesses are varied, while the total thickness is kept constant.

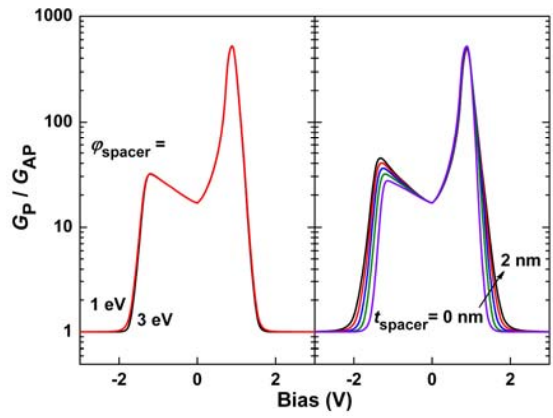

Fig.5 Bias dependence when the spacer layer varies in barrier height (left) and thickness (right). In the right figure the barrier thickness is varied from 0, 0.2, 0.5, 1, to 2 nm.